\documentclass[sigconf]{acmart}
\usepackage{multirow}

\AtBeginDocument{%
  }

\copyrightyear{2026}
\acmYear{2026}
\setcopyright{cc}
\setcctype{by}
\acmConference[RecSys '26]{20th ACM Conference on Recommender Systems}{September 27-October 02, 2026}{Minneapolis, MN, USA}
\acmBooktitle{20th ACM Conference on Recommender Systems (RecSys '26), September 27-October 02, 2026, Minneapolis, MN, USA}
\acmDOI{10.1145/3773078.3831853}
\acmISBN{979-8-4007-2284-4/2026/09}

\begin{document}

\title{Are We Really Making Progress in Group Recommendation? Unmasking the Tie-Breaking Illusion}


\author{Song-Duo Ma}
\orcid{0009-0006-2000-3921}
\affiliation{%
 \institution{National Taiwan University}
 \city{Taipei}
 \country{Taiwan}
}
\email{R14944001@ntu.edu.tw}

\author{Pu-Jen Cheng}
\orcid{0000-0001-5892-0385}
\affiliation{%
 \institution{National Taiwan University}
 \city{Taipei}
 \country{Taiwan}}
\email{pjcheng@csie.ntu.edu.tw}


\begin{abstract}
Recent group recommendation methods have reported strong improvements on standard benchmarks, but it remains unclear whether these gains always reflect genuine advances in modeling group preferences. In this paper, we show that several recent methods are affected by a systematic evaluation bias caused by the interaction between training-time score compression and evaluation-time deterministic tie-breaking. Specifically, an additional sigmoid transformation before the BPR objective can greatly increase tied top scores, making top-$K$ metrics such as HR@$K$ and NDCG@$K$ highly sensitive to how ties are resolved. We revisit recent representative methods and their baselines on \textsc{CAMRa2011} and \textsc{Mafengwo} under both group and user recommendation settings, and evaluate them with a tie-aware protocol that computes the exact expectation of HR@$K$ and NDCG@$K$ under uniform random tie-breaking. Our results show that many previously reported improvements shrink substantially under tie-aware evaluation, and the relative ranking of methods can change markedly. We further show that the additional sigmoid may act as implicit margin smoothing during optimization, and that temperature-scaled BPR can retain much of this benefit without inducing severe tie inflation. Overall, our findings highlight the importance of tie-aware evaluation for establishing reliable progress in group recommendation. The code is available at \url{https://github.com/songduoma/TieAwareGroupRec}.
\end{abstract}



\maketitle

\section{Introduction}

As online platforms and social media have become central to everyday information access and user decisions, recommender systems increasingly operate in scenarios where decisions are made not by a single user, but by a group, such as friends planning a trip or viewers selecting a movie together. In such settings, the system must infer a satisfactory collective choice from heterogeneous and often conflicting member preferences, making group recommendation substantially more challenging than standard single-user recommendation \citep{felfernig2018group,yin2019social,amer-yahia2009,baltrunas2010}. Motivated by this challenge, recent work has proposed increasingly sophisticated architectures to model group consensus, member interactions, and higher-order collaborative signals \citep{consrec2023, dhmae2024}.

However, progress in recommendation should be judged not only by new model designs, but also by the reliability of the evaluation protocols used to compare them. In recent group recommendation literature, methods are typically assessed with top-$K$ ranking metrics such as Hit Ratio at $K$ (HR@$K$) and Normalized Discounted Cumulative Gain at $K$ (NDCG@$K$) \citep{trattner2018evaluating, dinoia2022topn}. When evaluation procedures contain subtle implementation artifacts, reported improvements may not necessarily reflect genuine modeling advances.

This concern has been raised in prior recommender systems literature. Several studies have shown that some seemingly strong neural recommendation methods failed to maintain their claimed advantages under careful reproduction, rigorous comparison against strong baselines, or more careful hyperparameter tuning \citep{dacrema2019progress,dacrema2021troubling,everyonesawinner2023}. Inspired by this line of work, we revisit recent group recommendation methods and ask a simple question: are recent reported gains truly driven by better modeling of group preferences, or are they partly inflated by an overlooked evaluation artifact?

In this paper, we show that the issue is not merely due to incomplete reproduction or weak baseline comparison. Instead, it arises from an interaction between \emph{training-time score compression} and \emph{evaluation-time deterministic tie-breaking}. In several recent group recommendation codebases derived from or closely related to \textsc{ConsRec} \citep{consrec2023}, as well as in some baseline implementations distributed within these repositories, the predicted scores are passed through an additional sigmoid transformation before the Bayesian Personalized Ranking (BPR) \citep{rendle2009bpr} objective is applied. Although this design may appear innocuous, it can substantially compress score differences and increase the frequency of tied top scores, thereby producing many ties among candidate items at evaluation time. Once such ties become prevalent, top-$K$ ranking metrics can become highly sensitive to the tie-breaking rule, so the reported results may partly reflect implementation-dependent tie-breaking behavior rather than the model's actual ranking quality \citep{mcsherry2008tiedscores,lin2019scoreties,saha2022ties}.

As a motivating observation, Table~\ref{tab:intro_first_vs_last} shows what happens when the same score predictions are evaluated under two diagnostic tie-breaking rules on \textsc{CAMRa2011} group recommendation. In one case, the held-out positive item is placed at the earliest feasible position within its tied block; in the other, it is placed at the latest feasible position. Since both variants use exactly the same trained model and exactly the same prediction scores, any difference between them is caused solely by how ties are resolved. The resulting gaps are dramatic: for several recent implementations, the drops can exceed 90\% on multiple metrics, and in some cases the scores collapse to nearly zero. This suggests that tied scores are not a minor implementation detail, but can fundamentally alter empirical conclusions.

\begin{table}[t]
\centering
\caption{Motivating observation on \textsc{CAMRa2011} group recommendation. ``First'' and ``Last'' place the held-out positive item at the first or last feasible position within its tied block, respectively. The same score predictions can yield drastically different top-$K$ results solely due to tie-breaking.}

\label{tab:intro_first_vs_last}
\resizebox{\columnwidth}{!}{
\begin{tabular}{l|rrr|rrr|rrr}
\toprule
\multirow{2}{*}{Metric}
& \multicolumn{3}{c|}{\textsc{AlignGroup} (CIKM 2024)}
& \multicolumn{3}{c|}{\textsc{DHMAE} (SIGIR 2024)}
& \multicolumn{3}{c}{\textsc{ITR} (NeurIPS 2024)} \\
\cmidrule(lr){2-4}
\cmidrule(lr){5-7}
\cmidrule(l){8-10}
& First & Last & \% Change
& First & Last & \% Change
& First & Last & \% Change \\
\midrule
HR@1
& 0.7890 & 0.0028 & -99.65\%
& 0.9717 & 0.0000 & -100.00\%
& 0.6517 & 0.0248 & -96.19\% \\
HR@5
& 0.7993 & 0.1338 & -83.26\%
& 0.9717 & 0.0000 & -100.00\%
& 0.7007 & 0.2786 & -60.24\% \\
NDCG@5
& 0.7934 & 0.0600 & -92.43\%
& 0.9717 & 0.0000 & -100.00\%
& 0.6738 & 0.1384 & -79.47\% \\
\bottomrule
\end{tabular}
}
\end{table}

To study this issue, we revisit a family of recent group recommendation methods that report strong gains on standard benchmarks, including \textsc{ConsRec} \citep{consrec2023}, \textsc{AlignGroup} \citep{aligngroup2024}, \textsc{DHMAE} \citep{dhmae2024}, \textsc{ITR} \citep{itr2024}, and \textsc{DGGVAE} \citep{dggvae2026}, together with the baseline methods used in their original comparisons. We study both the \emph{group recommendation} setting and the associated \emph{user recommendation} setting on two widely used benchmarks, \textsc{CAMRa2011} and \textsc{Mafengwo} \citep{cao2018attentive}. We re-evaluate each method using both the original evaluation protocol adopted in prior codebases and our tie-aware evaluation protocol, which computes the exact expected HR@$K$ and NDCG@$K$ under uniform random tie-breaking. By applying both protocols to the same trained models and the same score predictions, our analysis allows us to separate genuine ranking quality from metric inflation induced by tied top scores and deterministic tie-breaking.

Our results reveal a systematic evaluation bias. Across multiple methods, tasks, and datasets, the original evaluation protocol can substantially overestimate performance when top-score ties are prevalent. This inflation is strongly associated with the prevalence and severity of top-score ties and can dramatically reshape the empirical ranking of recent methods. These findings suggest that part of the apparent progress in group recommendation may be an artifact of deterministic tie-breaking under score compression rather than genuine advances in modeling group preferences. This calls for a re-examination of recent methods under a more principled evaluation protocol.

Beyond re-evaluating recent methods, we further investigate why some implementations introduce an additional sigmoid transformation before the BPR objective. Our analysis suggests that, although this design induces severe tie inflation at evaluation time, it may also provide implicit margin smoothing during optimization. To disentangle these two effects, we examine temperature-scaled BPR as a simple alternative that smooths pairwise score differences directly without severely distorting the relative score ordering.

We summarize our main contributions as follows:
\begin{itemize}
    \item We uncover a systematic evaluation bias in recent group recommendation methods caused by the interaction between score compression and deterministic tie-breaking during top-$K$ evaluation.
    \item We propose a tie-aware evaluation protocol for HR@$K$ and NDCG@$K$ that computes their exact expectation under uniform random tie-breaking.
    \item Through a controlled re-evaluation of recent group recommendation methods and the baselines used in their original comparisons, we show that many reported improvements diminish substantially under tie-aware evaluation and that the relative ranking of methods can change markedly.
    \item We further interpret the additional sigmoid transformation before the BPR objective as an implicit form of margin smoothing, and examine temperature-scaled BPR as a simple alternative that preserves this effect without inducing severe tie inflation.
\end{itemize}

Overall, our results show that reliable progress in group recommendation requires reliable evaluation. By exposing the tie-breaking illusion behind several recent results, our study highlights the importance of reporting ranking performance together with tie-aware estimates and tie statistics when tied scores are prevalent.

\section{Preliminaries}

This section introduces the ranking formulation and evaluation concepts used throughout our study. We first define the group recommendation and user recommendation tasks under the standard top-$K$ evaluation protocol. We then describe the additional score transformation before BPR found in several recent implementations, and discuss how tied scores interact with deterministic tie-breaking to affect ranking evaluation.

\subsection{Task Definition and Standard Evaluation}
\label{sec:task_eval}

We study two closely related ranking settings commonly considered in recent group recommendation research: \emph{group recommendation} and \emph{user recommendation}. In the group recommendation setting, a model ranks candidate items for a group based on the group's collective preference. In the user recommendation setting, the model instead ranks candidate items for an individual user based on that user's historical interactions. We consider both settings because recent group recommendation methods commonly report results for both under the same ranking-based evaluation protocol.

Formally, let $x$ denote a test instance, representing either a group or an individual user depending on the setting. Each test instance is associated with an evaluation candidate set $\mathcal{I}_x$, and the recommendation model assigns a real-valued score $\hat{y}_{x,i}$ to each candidate item $i \in \mathcal{I}_x$. In the standard evaluation pipeline, candidate items are ordered by descending score, and the rank of the held-out positive item is then used to compute the evaluation metrics.

Following prior work, performance is measured using top-$K$ ranking metrics, primarily HR@$K$ and NDCG@$K$. Let $r_x$ denote the rank position of the held-out positive item for $x$. Since each test instance contains exactly one positive item, the instance-level metrics can be written as
\begin{equation}
\mathrm{HR@}K(x) = \mathbb{I}[r_x \le K],
\end{equation}
and
\begin{equation}
\mathrm{NDCG@}K(x) = \frac{\mathbb{I}[r_x \le K]}{\log_2(r_x + 1)},
\end{equation}
where $\mathbb{I}[\cdot]$ is the indicator function. The final reported score is obtained by averaging these quantities over all test instances.

These definitions are straightforward when the model induces a unique ordering over candidate items. When tied scores occur, the rank $r_x$ is no longer uniquely defined and instead depends on how ties are resolved. As a result, HR@$K$ and NDCG@$K$ can vary substantially across different tie-breaking rules.

\subsection{Score Transformation Before BPR}
\label{sec:pre_bpr_sigmoid}

Many recommendation models are trained with pairwise ranking objectives based on BPR \citep{rendle2009bpr}. For a training instance, which we also denote by $x$ for simplicity, let $i^+$ denote an observed positive item and $i^-$ denote a sampled negative item. A model first produces real-valued scores $s_{x,i^+}$ and $s_{x,i^-}$ for the two items, and BPR encourages the positive item to receive a higher score than the negative one by optimizing
\begin{equation}
\mathcal{L}_{\mathrm{BPR}}
= - \log \sigma \bigl(s_{x,i^+} - s_{x,i^-}\bigr),
\end{equation}
where $\sigma(\cdot)$ is the sigmoid function. Since the loss depends on the score difference $s_{x,i^+} - s_{x,i^-}$, standard BPR is typically applied directly to raw model scores.

However, in the family of recent group recommendation methods considered in this work, the training pipeline applies an additional sigmoid transformation to each item score before the BPR objective is computed. Concretely, instead of directly optimizing the raw scores $s_{x,i}$, these methods first define transformed scores
\begin{equation}
\tilde{s}_{x,i} = \sigma(s_{x,i}),
\end{equation}
and then optimize the pairwise objective
\begin{equation}
\mathcal{L}_{\mathrm{pre\mbox{-}BPR\ sigmoid}}
= - \log \sigma \bigl(\tilde{s}_{x,i^+} - \tilde{s}_{x,i^-}\bigr)
= - \log \sigma \bigl(\sigma(s_{x,i^+}) - \sigma(s_{x,i^-})\bigr).
\end{equation}
This differs from standard BPR, which already applies a sigmoid to the pairwise score difference. The additional transformation therefore introduces a second nonlinearity at the level of individual item scores, before the pairwise margin is formed.

This design has an important effect on score geometry. Because the sigmoid maps all real-valued inputs into the bounded interval $(0,1)$, it compresses the dynamic range of scores and shrinks pairwise score differences, particularly when raw scores lie in the saturated regions. Consequently, items with noticeably different raw scores may become much less separable after transformation. 

Although the sigmoid is monotonic and thus preserves pairwise order in exact arithmetic, this bounded transformation can flatten the score landscape near the top of the ranking in practice. In the models we study, it is associated with a marked increase in exact top-score ties at evaluation time, making top-$K$ evaluation highly sensitive to tie-breaking behavior.

\subsection{Ties and Deterministic Tie-Breaking}
\label{sec:ties_deterministic}

When multiple candidate items receive the same predicted score, the ranking of these items is no longer uniquely determined by the model outputs alone. In this case, the rank of the held-out positive item depends on an additional tie-breaking rule imposed by the evaluation pipeline. This issue is especially important for top-$K$ metrics because even a small change in tie-breaking rules can move the positive item across the top-$K$ cutoff and thus substantially affect the reported metric value. More specifically, for single-positive top-$K$ evaluation, the ambiguity is determined by the tied-block containing the held-out positive item. Whenever that block has size greater than one, the positive item does not have a unique rank, and its contribution to HR@$K$ and NDCG@$K$ may vary depending on how the tie is resolved.

While tied scores can occur at any position in the ranking, we pay particular attention to the special case of \emph{top-score ties}. Let
\begin{equation}
\mathcal{T}_x = \left\{ i \in \mathcal{I}_x : \hat{y}_{x,i} = \max_{j \in \mathcal{I}_x} \hat{y}_{x,j} \right\}
\end{equation}
denote the set of top-scoring items, and let
\begin{equation}
m_x = |\mathcal{T}_x|
\end{equation}
be the corresponding top-score tie size. When $m_x >1$, we say that instance $x$ contains a \emph{top-score tie}. We emphasize top-score ties for two reasons. First, they have the strongest potential effect on the top of the ranked list, where HR@$K$ and NDCG@$K$ are most sensitive to rank changes. Second, they provide a simple empirical signature of score compression in the methods studied here.

In principle, ties may be resolved in several ways. One may break ties uniformly at random, or use a deterministic secondary key such as the original candidate order. Prior work has shown that deterministic tie-breaking can introduce bias into ranking evaluation and undermine the repeatability of empirical comparisons when tied scores are common \citep{cabanac2010tiebreaking,lin2019scoreties}.

This is exactly the evaluation pattern implemented in the family of codebases examined in this work. During evaluation, each test instance is constructed by placing the held-out positive item first in the candidate list, followed by sampled negative items. The candidate scores are then sorted in descending order by a deterministic ranking routine. As a result, whenever the positive item is tied with one or more negatives, its final rank can be influenced by the original candidate order rather than by the model scores alone. In the examined implementations, this ordering tends to assign the positive item the most favorable feasible position within its tied block, thereby yielding an optimistic estimate of ranking quality whenever such ties are frequent.

This observation motivates a tie-aware evaluation protocol that accounts explicitly for the uncertainty induced by tied scores, rather than relying on a single deterministic tie-breaking rule.

\section{Tie-Aware Evaluation Framework}

This section presents the tie-aware evaluation framework used in our empirical analysis. We first define tie-aware HR@$K$ and NDCG@$K$, which account for rank ambiguity under tied scores by computing the exact expected metric under uniform random tie-breaking. We then describe how this framework is applied in our experiments, including the datasets, reproduced methods, candidate-set protocols, evaluation variants, and tie statistics.

\subsection{Tie-Aware Metrics}
\label{sec:tie_aware_metrics}

To avoid the optimistic bias induced by deterministic tie-breaking, we adopt a tie-aware evaluation that accounts for the rank ambiguity caused by tied scores. Rather than assigning the positive item a single deterministic rank, we evaluate its expected contribution to HR@$K$ and NDCG@$K$ under uniform random tie-breaking.

In this work, we consider \emph{uniform random tie-breaking}: whenever the held-out positive item is tied with other candidates, its rank is assumed to be uniformly distributed over all feasible positions within its tied block. We then define the tie-aware evaluation metric as the exact expectation of the standard ranking metric under this tie-breaking process.

Formally, for a test instance $x$, let $i^+$ denote the held-out positive item. We define
\begin{equation}
a_x = \left| \left\{ j \in \mathcal{I}_x : \hat{y}_{x,j} > \hat{y}_{x,i^+} \right\} \right|
\end{equation}
as the number of candidate items whose predicted score is strictly higher than that of the positive item, and
\begin{equation}
b_x = \left| \left\{ j \in \mathcal{I}_x : \hat{y}_{x,j} = \hat{y}_{x,i^+} \right\} \right|
\end{equation}
as the size of the tied block containing the positive item, including the positive item itself. Under uniform random tie-breaking, the rank of the positive item is uniformly distributed over the set
\begin{equation}
\left\{ a_x + 1,\; a_x + 2,\; \dots,\; a_x + b_x \right\}.
\end{equation}

Accordingly, the tie-aware HR@$K$ for instance $x$ is the probability that the positive item falls within the top-$K$ positions:
\begin{equation}
\mathrm{HR@}K_{\mathrm{tie}}(x)
=
\frac{
\max \left(0,\; \min(K,\; a_x + b_x) - a_x \right)
}{b_x}.
\label{eq:tie_hr}
\end{equation}
Equivalently, this is the fraction of feasible ranks in the positive item's tied block that lie within the top-$K$ cutoff.

Similarly, the tie-aware NDCG@$K$ is defined as the exact expected discounted gain over all feasible ranks of the positive item:
\begin{equation}
\mathrm{NDCG@}K_{\mathrm{tie}}(x)
=
\frac{1}{b_x}
\sum_{r = a_x + 1}^{a_x + b_x}
\frac{\mathbb{I}[r \le K]}{\log_2(r+1)}.
\label{eq:tie_ndcg}
\end{equation}

The final tie-aware metrics are obtained by averaging the instance-level expectations over all test instances:
\begin{equation}
\mathrm{HR@}K_{\mathrm{tie}}
=
\frac{1}{|\mathcal{X}_{\mathrm{test}}|}
\sum_{x \in \mathcal{X}_{\mathrm{test}}}
\mathrm{HR@}K_{\mathrm{tie}}(x),
\end{equation}
and
\begin{equation}
\mathrm{NDCG@}K_{\mathrm{tie}}
=
\frac{1}{|\mathcal{X}_{\mathrm{test}}|}
\sum_{x \in \mathcal{X}_{\mathrm{test}}}
\mathrm{NDCG@}K_{\mathrm{tie}}(x).
\end{equation}

This protocol reduces to exactly the standard evaluation in Section~\ref{sec:task_eval} when no tie occurs for the positive item, i.e., when $b_x = 1$. When $b_x > 1$, the positive item does not have a unique rank, and a deterministic evaluation must implicitly choose one feasible position from its tied block. The proposed tie-aware metric instead averages over all feasible positions and can therefore be interpreted as the exact expected metric under uniform random tie-breaking.

In our experiments, we use this tie-aware protocol as the primary reference evaluation, which avoids relying on arbitrary deterministic ordering among items with identical scores.

\subsection{Evaluation Protocol}
\label{sec:eval_protocol}

We evaluate recent group recommendation methods on two widely used benchmark datasets, \textsc{CAMRa2011} and \textsc{Mafengwo} \citep{cao2018attentive}, under both the \emph{group recommendation} and \emph{user recommendation} settings. Table~\ref{tab:dataset_statistics} summarizes the basic statistics of the two datasets, including the numbers of users, items, groups, user--item interactions, and group--item interactions. For all methods, we use the original training and test splits provided in the released codebases and retrain every model from scratch. Unless otherwise noted, the main results are averaged over three random seeds.

\begin{table}[!t]
\footnotesize
\centering
\caption{Statistics of the two benchmark datasets.}
\label{tab:dataset_statistics}
\begin{tabular}{lrrrrr}
\toprule
Dataset & Users & Items & Groups & U-I Interactions & G-I Interactions \\
\midrule
\textsc{Mafengwo} & 5,275 & 1,513 & 995 & 39,761 & 3,595 \\
\textsc{CAMRa2011} & 602 & 7,710 & 290 & 116,344 & 145,068 \\
\bottomrule
\end{tabular}
\end{table}

We reproduce five representative recent methods: \textsc{ConsRec} \citep{consrec2023}\footnote{\url{https://github.com/FDUDSDE/WWW2023ConsRec}}, \textsc{AlignGroup} \citep{aligngroup2024}\footnote{\url{https://github.com/Jinfeng-Xu/AlignGroup}}, \textsc{DHMAE} \citep{dhmae2024}\footnote{\url{https://github.com/ICharlotteI/DHMAE}}, \textsc{ITR} \citep{itr2024}\footnote{\url{https://github.com/yueliu1999/ITR}}, and \textsc{DGGVAE} \citep{dggvae2026}\footnote{\url{https://github.com/Jinfeng-Xu/DGGVAE}}, using their official released implementations. We also reproduce the baseline methods using the implementations released by the \textsc{ConsRec} authors\footnote{\url{https://github.com/FDUDSDE/WWW2023GroupRecBaselines}}. We follow the original training and evaluation protocols of each method as closely as possible. In particular, \textsc{ConsRec}, \textsc{AlignGroup}, \textsc{ITR}, and \textsc{DGGVAE} adopt a single-positive sampled evaluation setting, in which each test instance consists of one held-out positive item and 100 sampled negative items. In contrast, \textsc{DHMAE} follows its original all-item ranking protocol, in which the held-out positive item is ranked against the full candidate item set. Because these evaluation settings differ substantially in candidate set size and ranking difficulty, we preserve each method's original evaluation protocol in our main reproduction. We therefore interpret the results primarily as comparisons across evaluation variants in the same method, rather than as directly comparable scores across different evaluation settings.

To ensure that differences between evaluation protocols are not confounded by changes in model outputs, we compute all evaluation results from the \emph{same} score predictions of a trained model. For each random seed, trained checkpoint, and test instance, candidate scores are computed once and then reused to obtain two evaluation results: (1) the original evaluation protocol implemented in the released codebase, and (2) the proposed \emph{tie-aware} evaluation defined in Section~\ref{sec:tie_aware_metrics}. We then average the resulting metrics over three random seeds. This design isolates evaluation-side effects from retraining variability within each seed, since both evaluation results are computed from identical model outputs.

In the released codebases we examined, candidate scores are ranked using \texttt{np.argsort} from NumPy \citep{harris2020array} with its default sorting algorithm. We treat this as the \emph{original} evaluation protocol used in the released implementations. Since this protocol assigns a single deterministic rank to each candidate item, its output can depend on implementation-specific ordering behavior when tied scores occur. The proposed tie-aware evaluation instead computes the expected contribution of the held-out positive item over all feasible positions within its tied block, thereby avoiding reliance on a particular deterministic ordering among tied items.

\begin{table*}[t]
\centering
\caption{Performance inflation under the original evaluation protocol. Results are averaged over three random seeds. For each metric, we report the original score, the tie-aware score, and the relative change from original to tie-aware evaluation. Larger negative changes indicate greater overestimation caused by deterministic tie-breaking under tied scores.}
\label{tab:main_original_vs_tieaware}
\resizebox{\textwidth}{!}{
\begin{tabular}{ll|ccr|ccr|ccr|ccr|ccr}
\toprule
\multirow{2}{*}{Dataset} & \multirow{2}{*}{Metric}
& \multicolumn{3}{c|}{\textsc{ConsRec} (WWW 2023)}
& \multicolumn{3}{c|}{\textsc{AlignGroup} (CIKM 2024)}
& \multicolumn{3}{c|}{\textsc{DHMAE}\textsuperscript{\textdagger} (SIGIR 2024)}
& \multicolumn{3}{c|}{\textsc{ITR} (NeurIPS 2024)}
& \multicolumn{3}{c}{\textsc{DGGVAE} (TOIS 2026)} \\
\cmidrule(lr){3-5} \cmidrule(lr){6-8} \cmidrule(lr){9-11} \cmidrule(lr){12-14} \cmidrule(l){15-17}
&
& Original & Tie-aware & \% Change
& Original & Tie-aware & \% Change
& Original & Tie-aware & \% Change
& Original & Tie-aware & \% Change
& Original & Tie-aware & \% Change \\
\midrule

\multicolumn{17}{l}{\textbf{Group Recommendation Task}} \\
\midrule
\multirow{5}{*}{CAMRa2011}
& HR@1    & 0.2228 & 0.2209 & -0.82\% & 0.7400 & 0.1094 & -85.22\% & 0.9782 & 0.0002 & -99.98\% & 0.6057 & 0.1420 & -76.56\% & 0.9320 & 0.0611 & -93.45\% \\
& HR@5    & 0.6343 & 0.6339 & -0.05\% & 0.7657 & 0.5091 & -33.52\% & 0.9782 & 0.0009 & -99.90\% & 0.6857 & 0.5714 & -16.68\% & 0.9320 & 0.3054 & -67.23\% \\
& HR@10   & 0.8235 & 0.8236 & 0.01\%  & 0.8324 & 0.7776 & -6.58\%  & 0.9782 & 0.0019 & -99.81\% & 0.8145 & 0.8079 & -0.81\%  & 0.9326 & 0.5913 & -36.60\% \\
& NDCG@5  & 0.4315 & 0.4305 & -0.23\% & 0.7514 & 0.3065 & -59.21\% & 0.9782 & 0.0005 & -99.94\% & 0.6424 & 0.3554 & -44.68\% & 0.9320 & 0.1801 & -80.68\% \\
& NDCG@10 & 0.4931 & 0.4923 & -0.17\% & 0.7726 & 0.3939 & -49.01\% & 0.9782 & 0.0008 & -99.91\% & 0.6838 & 0.4328 & -36.71\% & 0.9322 & 0.2716 & -70.86\% \\
\midrule
\multirow{5}{*}{Mafengwo}
& HR@1    & 0.6328 & 0.6152 & -2.79\% & 0.7357 & 0.5007 & -31.94\% & 0.9454 & 0.0018 & -99.81\% & 0.8144 & 0.0084 & -98.97\% & 0.7119 & 0.5558 & -21.93\% \\
& HR@5    & 0.8807 & 0.8677 & -1.48\% & 0.9716 & 0.6525 & -32.84\% & 0.9454 & 0.0091 & -99.04\% & 0.9384 & 0.0473 & -94.96\% & 0.8596 & 0.8570 & -0.31\% \\
& HR@10   & 0.9116 & 0.8907 & -2.29\% & 0.9739 & 0.6709 & -31.11\% & 0.9454 & 0.0182 & -98.07\% & 0.9441 & 0.0974 & -89.69\% & 0.8908 & 0.8875 & -0.37\% \\
& NDCG@5  & 0.7710 & 0.7575 & -1.75\% & 0.8763 & 0.5903 & -32.63\% & 0.9454 & 0.0054 & -99.43\% & 0.8849 & 0.0273 & -96.92\% & 0.7916 & 0.7261 & -8.28\% \\
& NDCG@10 & 0.7810 & 0.7650 & -2.06\% & 0.8770 & 0.5962 & -32.02\% & 0.9454 & 0.0083 & -99.12\% & 0.8868 & 0.0433 & -95.12\% & 0.8018 & 0.7361 & -8.19\% \\
\midrule

\multicolumn{17}{l}{\textbf{User Recommendation Task}} \\
\midrule
\multirow{5}{*}{CAMRa2011}
& HR@1    & 0.2219 & 0.2201 & -0.82\% & 0.7658 & 0.1036 & -86.47\% & 0.2997 & 0.0013 & -99.56\% & 0.6105 & 0.1492 & -75.57\% & 0.5906 & 0.1550 & -73.76\% \\
& HR@5    & 0.6730 & 0.6726 & -0.05\% & 0.7831 & 0.4890 & -37.55\% & 0.3013 & 0.0070 & -97.68\% & 0.7165 & 0.6047 & -15.60\% & 0.6992 & 0.6206 & -11.24\% \\
& HR@10   & 0.8421 & 0.8422 & 0.02\%  & 0.8396 & 0.7620 & -9.24\%  & 0.3025 & 0.0142 & -95.31\% & 0.8306 & 0.8210 & -1.15\%  & 0.8247 & 0.8178 & -0.83\% \\
& NDCG@5  & 0.4530 & 0.4520 & -0.21\% & 0.7731 & 0.2932 & -62.08\% & 0.3006 & 0.0041 & -98.63\% & 0.6614 & 0.3780 & -42.86\% & 0.6404 & 0.3897 & -39.15\% \\
& NDCG@10 & 0.5084 & 0.5075 & -0.17\% & 0.7912 & 0.3820 & -51.72\% & 0.3010 & 0.0064 & -97.86\% & 0.6984 & 0.4487 & -35.75\% & 0.6810 & 0.4542 & -33.29\% \\
\midrule
\multirow{5}{*}{Mafengwo}
& HR@1    & 0.5967 & 0.2408 & -59.64\% & 0.6805 & 0.2343 & -65.57\% & 0.6254 & 0.0607 & -90.30\% & 0.6361 & 0.2189 & -65.59\% & 0.5407 & 0.2347 & -56.60\% \\
& HR@5    & 0.7727 & 0.6512 & -15.73\% & 0.8103 & 0.6473 & -20.12\% & 0.6346 & 0.2922 & -53.94\% & 0.7832 & 0.6322 & -19.27\% & 0.7372 & 0.6308 & -14.42\% \\
& HR@10   & 0.8385 & 0.7752 & -7.55\%  & 0.8636 & 0.7738 & -10.39\% & 0.6712 & 0.5253 & -21.75\% & 0.8379 & 0.7465 & -10.91\% & 0.8136 & 0.7502 & -7.79\% \\
& NDCG@5  & 0.6861 & 0.4582 & -33.21\% & 0.7461 & 0.4520 & -39.42\% & 0.6295 & 0.1741 & -72.35\% & 0.7097 & 0.4365 & -38.49\% & 0.6426 & 0.4442 & -30.88\% \\
& NDCG@10 & 0.7077 & 0.4986 & -29.55\% & 0.7637 & 0.4932 & -35.42\% & 0.6410 & 0.2488 & -61.18\% & 0.7277 & 0.4737 & -34.90\% & 0.6675 & 0.4835 & -27.56\% \\
\bottomrule
\end{tabular}
}

\vspace{2pt}
\begin{minipage}{\textwidth}
\footnotesize
\textsuperscript{\textdagger}\textsc{DHMAE} follows its original all-item evaluation protocol, whereas the other methods use the sampled evaluation protocol with one positive item and 100 negative items. Therefore, absolute scores involving \textsc{DHMAE} are not directly comparable with those of the other methods; the intended comparison is between evaluation variants within the same method.
\end{minipage}

\end{table*}

We report HR@$K$ and NDCG@$K$ under both the original and tie-aware evaluation protocols, averaging metrics over three random seeds for the main evaluation. To quantify the prevalence of tied top scores, we additionally report several tie statistics based on the top-score tie size $m_x$ defined in Section~\ref{sec:ties_deterministic}. Specifically, we report the average top-score tie size $\mathbb{E}[m_x]$, the average top-score tie ratio $\mathbb{E}[m_x / |\mathcal{I}_x|]$, and the proportion of test instances with $m_x > 1$. Throughout our experiments, ties are defined by exact score equality, without any numerical tolerance. These statistics complement HR@$K$ and NDCG@$K$ by quantifying the prevalence of top-score ties and helping interpret how much the reported ranking performance can depend on tie-breaking behavior.

Following the released implementations, all reproduced experiments use single-precision (FP32) computation without automatic mixed precision. Predicted scores are transferred directly from FP32 PyTorch tensors to NumPy arrays for evaluation. We note that numerical precision can interact with score compression: although the sigmoid is strictly monotonic in exact arithmetic, sufficiently saturated scores may collapse to identical values under finite-precision computation.

\section{Main Findings}

In this section, we present the main empirical findings of our study. We first show that the original evaluation protocol can substantially overestimate performance relative to tie-aware evaluation. We then examine how these gaps relate to top-score tie prevalence, and finally revisit the relative performance of methods after removing the additional sigmoid from the affected implementations.

\subsection{Inflation Under the Original Protocol}
\label{sec:main_findings_inflation}

We first compare the performance produced by the original evaluation protocol in the released codebases with those obtained under our tie-aware evaluation protocol. Table~\ref{tab:main_original_vs_tieaware} summarizes the results for the five representative methods introduced in Section~\ref{sec:eval_protocol}, covering both the \emph{group recommendation} and \emph{user recommendation} tasks. For each method and metric, we report the mean original score, the mean tie-aware score, and the relative percentage change computed from these mean values.

Across both tasks, the original protocol generally produces more optimistic results than tie-aware evaluation, but the magnitude of this inflation varies substantially across methods. Among the five methods, \textsc{ConsRec} is the least sensitive overall: its original and tie-aware scores remain close in most settings, especially on \textsc{CAMRa2011}. Although the gaps become more visible on \textsc{Mafengwo}, \textsc{ConsRec} is still comparatively less affected than the other methods.

By contrast, the remaining methods often experience dramatic drops under tie-aware evaluation. For example, in the \emph{group recommendation} task on \textsc{CAMRa2011}, \textsc{DHMAE} drops from 0.9782 to 0.0002 on HR@1, while \textsc{DGGVAE} drops from 0.9320 to 0.0611. On \textsc{Mafengwo}, \textsc{ITR} falls from 0.8144 to 0.0084 on HR@1. A similar pattern is observed in the \emph{user recommendation} task, where several methods show substantial declines across both datasets. These discrepancies are especially pronounced at smaller cutoff values such as HR@1, where the contribution of the positive item depends most strongly on how ties are resolved.

Overall, these results provide initial evidence that the original deterministic evaluation protocol can substantially overestimate the performance of several recent group recommendation methods. More importantly, the extent of this inflation is large enough to reshape the relative ordering of methods, suggesting that some reported gains in recent group recommendation literature may be driven by evaluation artifacts rather than genuine improvements in ranking quality.

\subsection{Inflation Scales with Top-Score Ties}
\label{sec:main_findings_tie_stats}

\begin{table}[t]
\centering
\caption{Top-score tie statistics for the five representative methods, averaged over three random seeds. Larger values indicate more prevalent or severe top-score ties.}
\label{tab:tie_statistics}
\resizebox{\columnwidth}{!}{
\begin{tabular}{ll|ccccc}
\toprule
Dataset & Statistic
& \textsc{ConsRec}
& \textsc{AlignGroup}
& \textsc{DHMAE}
& \textsc{ITR}
& \textsc{DGGVAE} \\
\midrule
\multicolumn{7}{l}{\textbf{Group Recommendation Task}} \\
\midrule
\multirow{3}{*}{CAMRa2011}
& Avg. top-score tie size
& 1.00 & 8.58 & 5524.47 & 5.31 & 18.17 \\
& Avg. top-score tie ratio
& 0.99\% & 8.49\% & 75.52\% & 5.25\% & 17.99\% \\
& Top-score tie frequency
& 0.37\% & 98.41\% & 100.00\% & 86.41\% & 100.00\% \\
\midrule
\multirow{3}{*}{Mafengwo}
& Avg. top-score tie size
& 1.18 & 23.74 & 637.81 & 82.40 & 1.55 \\
& Avg. top-score tie ratio
& 1.16\% & 23.50\% & 42.23\% & 81.58\% & 1.53\% \\
& Top-score tie frequency
& 4.15\% & 25.16\% & 100.00\% & 84.52\% & 33.57\% \\
\midrule
\multicolumn{7}{l}{\textbf{User Recommendation Task}} \\
\midrule
\multirow{3}{*}{CAMRa2011}
& Avg. top-score tie size
& 1.00 & 9.18 & 236.16 & 5.05 & 4.76 \\
& Avg. top-score tie ratio
& 0.99\% & 9.08\% & 3.17\% & 5.00\% & 4.71\% \\
& Top-score tie frequency
& 0.37\% & 98.95\% & 84.94\% & 81.27\% & 88.54\% \\
\midrule
\multirow{3}{*}{Mafengwo}
& Avg. top-score tie size
& 6.95 & 7.66 & 12.89 & 8.48 & 6.79 \\
& Avg. top-score tie ratio
& 6.88\% & 7.58\% & 0.87\% & 8.40\% & 6.73\% \\
& Top-score tie frequency
& 75.07\% & 79.86\% & 98.57\% & 81.55\% & 67.90\% \\
\bottomrule
\end{tabular}
}
\end{table}

\begin{figure*}[t]
    \centering
    \includegraphics[width=\textwidth]{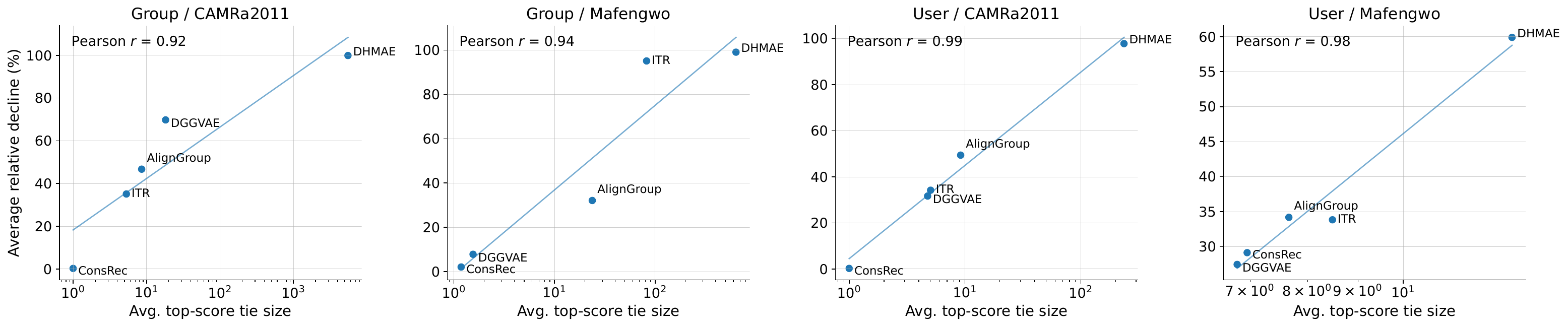}
    \caption{Average top-score tie size versus average relative metric drop from original to tie-aware evaluation, with both quantities averaged over three random seeds. Each point denotes one method under one task--dataset setting. Pearson correlations are computed using the log-transformed average top-score tie size, with $r$ ranging from 0.92 to 0.99 across settings.}
    \Description{Four scatter plots showing the relationship between average top-score tie size and the average relative performance decline from original to tie-aware evaluation for group and user recommendation on CAMRa2011 and Mafengwo. Larger tied blocks are consistently associated with larger performance declines, with Pearson correlations ranging from 0.92 to 0.99.}
    \label{fig:tie_count_vs_avg_drop}
\end{figure*}

Table~\ref{tab:tie_statistics} reports the top-score tie statistics averaged over three random seeds for all five methods, separately for the \emph{group recommendation} and \emph{user recommendation} tasks. Across both tasks, \textsc{ConsRec} consistently exhibits the lowest prevalence of top-score ties, whereas the other methods often produce substantially larger tied blocks near the top of the ranking.

In the \emph{group recommendation} task, the contrast is particularly striking. On \textsc{CAMRa2011}, \textsc{ConsRec} exhibits almost no top-score ties, whereas \textsc{DHMAE} produces extremely large tied blocks and top-score ties in all test instances. The remaining methods also show substantial tie prevalence. A similar pattern appears on \textsc{Mafengwo}, where top-score ties remain widespread for several methods.

The \emph{user recommendation} task shows the same overall trend, although the exact tie patterns differ by dataset. On \textsc{CAMRa2011}, \textsc{ConsRec} again remains largely unaffected, while several other methods exhibit top-score ties in most test instances. On \textsc{Mafengwo}, top-score ties become common across all methods, though their severity still varies substantially across models. Notably, \textsc{DHMAE} exhibits top-score ties in nearly all instances while still having a relatively small average tie ratio, partly because it follows an all-item ranking protocol with a much larger candidate set. This suggests that tie frequency, tie size, and tie ratio capture complementary aspects of score compression.

To quantify the relationship between top-score tie prevalence and metric
inflation, we compare the average top-score tie size with the average
relative drop across the five evaluation metrics reported in
Section~\ref{sec:main_findings_inflation}. Figure~\ref{fig:tie_count_vs_avg_drop} shows a clear positive association:
methods with larger top-score tied blocks consistently suffer larger
performance declines when moving from the original evaluation protocol
to tie-aware evaluation. Since tie sizes vary by several orders of
magnitude, we visualize them on a logarithmic scale and compute Pearson
correlations using the log-transformed tie size. Across settings, the
correlation coefficients range from 0.92 to 0.99.

Taken together, these results strongly support our central claim: performance inflation under the original evaluation protocol is systematically associated with the prevalence of top-score ties. As top-score ties become larger, the reported performance becomes increasingly sensitive to deterministic tie-breaking.

\begin{table*}[t]
\centering
\caption{
Tie-aware performance after removing the additional sigmoid before the BPR objective, averaged over three random seeds. For each method, we report the corrected tie-aware score, and for affected methods, the relative change from the original tie-aware score is shown in parentheses. The best and second-best corrected results among methods using the same sampled evaluation protocol are highlighted in bold and underline, respectively.
}
\label{tab:corrected_tieaware_results}
\resizebox{\textwidth}{!}{
\begin{tabular}{ll|cccccc|ccccc}
\toprule
\multirow{2}{*}{Dataset} & \multirow{2}{*}{Metric}
& \multicolumn{6}{c|}{\textbf{Baselines}}
& \multicolumn{5}{c}{\textbf{Recent Representative Methods}} \\
\cmidrule(lr){3-8} \cmidrule(l){9-13}
&
& \textsc{AGREE}
& \textsc{HyperGroup}
& \textsc{HCR}
& \textsc{GroupIM}
& \textsc{S$^2$-HHGR}
& \textsc{CubeRec}
& \textsc{ConsRec}
& \textsc{AlignGroup}
& \textsc{DHMAE}\textsuperscript{\textdagger}
& \textsc{ITR}
& \textsc{DGGVAE} \\
\midrule
\multicolumn{13}{l}{\textbf{Group Recommendation Task}} \\
\midrule
\multirow{5}{*}{CAMRa2011}
& HR@1
& 0.1724 {\footnotesize (-7.42\%)}
& 0.2128 {\footnotesize (+17.45\%)}
& 0.2022 {\footnotesize (+10.60\%)}
& 0.2208
& 0.1644
& \underline{0.2248}
& 0.2209
& 0.2153 {\footnotesize (+96.78\%)}
& 0.0002 {\footnotesize (+23.21\%)}
& 0.2128 {\footnotesize (+49.86\%)}
& \textbf{0.2269} {\footnotesize (+271.49\%)} \\
& HR@5
& 0.6098 {\footnotesize (+0.66\%)}
& 0.6256 {\footnotesize (+6.33\%)}
& 0.5933 {\footnotesize (-0.97\%)}
& 0.6203
& 0.6143
& \textbf{0.6438}
& 0.6339
& 0.6137 {\footnotesize (+20.54\%)}
& 0.0005 {\footnotesize (-50.71\%)}
& 0.6215 {\footnotesize (+8.77\%)}
& \underline{0.6408} {\footnotesize (+109.84\%)} \\
& HR@10
& 0.8121 {\footnotesize (+0.33\%)}
& 0.8076 {\footnotesize (+1.14\%)}
& 0.7914 {\footnotesize (-1.23\%)}
& 0.8066
& 0.8210
& \textbf{0.8345}
& \underline{0.8236}
& 0.7987 {\footnotesize (+2.72\%)}
& 0.0025 {\footnotesize (+36.02\%)}
& 0.8098 {\footnotesize (+0.23\%)}
& 0.8230 {\footnotesize (+39.18\%)} \\
& NDCG@5
& 0.3910 {\footnotesize (-2.25\%)}
& 0.4241 {\footnotesize (+9.84\%)}
& 0.4017 {\footnotesize (+1.81\%)}
& 0.4234
& 0.3927
& \textbf{0.4376}
& 0.4305
& 0.4173 {\footnotesize (+36.14\%)}
& 0.0003 {\footnotesize (-37.20\%)}
& 0.4196 {\footnotesize (+18.07\%)}
& \underline{0.4342} {\footnotesize (+141.12\%)} \\
& NDCG@10
& 0.4607 {\footnotesize (-1.27\%)}
& 0.4835 {\footnotesize (+6.34\%)}
& 0.4660 {\footnotesize (+1.21\%)}
& 0.4842
& 0.4603
& \textbf{0.5001}
& 0.4923
& 0.4777 {\footnotesize (+21.28\%)}
& 0.0010 {\footnotesize (+16.54\%)}
& 0.4810 {\footnotesize (+11.14\%)}
& \underline{0.4936} {\footnotesize (+81.74\%)} \\
\midrule
\multirow{5}{*}{Mafengwo}
& HR@1
& 0.4831 {\footnotesize (+10.01\%)}
& 0.4921 {\footnotesize (+30.00\%)}
& 0.4823 {\footnotesize (+6.82\%)}
& 0.3263
& 0.0186
& \textbf{0.6407}
& 0.4901 {\footnotesize (-20.33\%)}
& 0.6106 {\footnotesize (+21.94\%)}
& 0.0871 {\footnotesize (+4676.97\%)}
& 0.0040 {\footnotesize (-52.16\%)}
& \underline{0.6333} {\footnotesize (+13.96\%)} \\
& HR@5
& 0.7611 {\footnotesize (+10.60\%)}
& 0.7776 {\footnotesize (+13.69\%)}
& 0.7791 {\footnotesize (+2.95\%)}
& 0.6409
& 0.0680
& \underline{0.8653}
& 0.7531 {\footnotesize (-13.20\%)}
& 0.8506 {\footnotesize (+30.36\%)}
& 0.2161 {\footnotesize (+2271.03\%)}
& 0.0141 {\footnotesize (-70.22\%)}
& \textbf{0.8666} {\footnotesize (+1.12\%)} \\
& HR@10
& 0.8342 {\footnotesize (+12.16\%)}
& 0.8384 {\footnotesize (+7.79\%)}
& 0.8564 {\footnotesize (+4.88\%)}
& 0.7347
& 0.1169
& \textbf{0.9010}
& 0.8270 {\footnotesize (-7.16\%)}
& 0.8915 {\footnotesize (+32.87\%)}
& 0.3692 {\footnotesize (+1925.85\%)}
& 0.0290 {\footnotesize (-70.23\%)}
& \underline{0.8955} {\footnotesize (+0.90\%)} \\
& NDCG@5
& 0.6347 {\footnotesize (+10.62\%)}
& 0.6516 {\footnotesize (+19.93\%)}
& 0.6436 {\footnotesize (+4.10\%)}
& 0.4937
& 0.0434
& \textbf{0.7672}
& 0.6264 {\footnotesize (-17.31\%)}
& 0.7458 {\footnotesize (+26.33\%)}
& 0.1525 {\footnotesize (+2737.72\%)}
& 0.0087 {\footnotesize (-68.22\%)}
& \underline{0.7661} {\footnotesize (+5.50\%)} \\
& NDCG@10
& 0.6590 {\footnotesize (+11.33\%)}
& 0.6714 {\footnotesize (+17.04\%)}
& 0.6689 {\footnotesize (+4.90\%)}
& 0.5241
& 0.0589
& \textbf{0.7788}
& 0.6570 {\footnotesize (-14.12\%)}
& 0.7592 {\footnotesize (+27.33\%)}
& 0.2011 {\footnotesize (+2329.07\%)}
& 0.0134 {\footnotesize (-69.08\%)}
& \underline{0.7756} {\footnotesize (+5.37\%)} \\
\midrule
\multicolumn{13}{l}{\textbf{User Recommendation Task}} \\
\midrule
\multirow{5}{*}{CAMRa2011}
& HR@1
& 0.1844 {\footnotesize (-2.27\%)}
& 0.2096 {\footnotesize (+102.04\%)}
& 0.1915 {\footnotesize (+1.46\%)}
& 0.2049
& 0.1787
& 0.1515
& \textbf{0.2201}
& \underline{0.2199} {\footnotesize (+112.20\%)}
& 0.0092 {\footnotesize (+589.25\%)}
& 0.1992 {\footnotesize (+33.56\%)}
& 0.2029 {\footnotesize (+30.96\%)} \\
& HR@5
& \textbf{0.6743} {\footnotesize (+4.10\%)}
& 0.6627 {\footnotesize (+33.81\%)}
& 0.6310 {\footnotesize (-2.06\%)}
& 0.6269
& 0.6352
& 0.5360
& \underline{0.6726}
& 0.6672 {\footnotesize (+36.44\%)}
& 0.0410 {\footnotesize (+485.05\%)}
& 0.6449 {\footnotesize (+6.63\%)}
& 0.6557 {\footnotesize (+5.66\%)} \\
& HR@10
& \textbf{0.8519} {\footnotesize (+2.27\%)}
& 0.8382 {\footnotesize (+7.44\%)}
& 0.8026 {\footnotesize (-2.94\%)}
& 0.8023
& 0.8253
& 0.7598
& \underline{0.8422}
& 0.8328 {\footnotesize (+9.28\%)}
& 0.0750 {\footnotesize (+428.23\%)}
& 0.8226 {\footnotesize (+0.20\%)}
& 0.8189 {\footnotesize (+0.13\%)} \\
& NDCG@5
& 0.4353 {\footnotesize (+2.81\%)}
& 0.4425 {\footnotesize (+49.47\%)}
& 0.4179 {\footnotesize (-0.89\%)}
& 0.4222
& 0.4017
& 0.3475
& \textbf{0.4520}
& \underline{0.4489} {\footnotesize (+53.11\%)}
& 0.0248 {\footnotesize (+501.13\%)}
& 0.4294 {\footnotesize (+13.60\%)}
& 0.4347 {\footnotesize (+11.56\%)} \\
& NDCG@10
& 0.4935 {\footnotesize (+1.96\%)}
& 0.4998 {\footnotesize (+28.62\%)}
& 0.4740 {\footnotesize (-1.53\%)}
& 0.4793
& 0.4677
& 0.4201
& \textbf{0.5075}
& \underline{0.5031} {\footnotesize (+31.71\%)}
& 0.0356 {\footnotesize (+454.30\%)}
& 0.4873 {\footnotesize (+8.61\%)}
& 0.4880 {\footnotesize (+7.43\%)} \\
\midrule
\multirow{5}{*}{Mafengwo}
& HR@1
& 0.2308 {\footnotesize (-12.21\%)}
& 0.2222 {\footnotesize (-10.58\%)}
& 0.2504 {\footnotesize (+8.43\%)}
& 0.2259
& 0.2393
& 0.0421
& \underline{0.2532} {\footnotesize (+5.12\%)}
& \textbf{0.2717} {\footnotesize (+15.97\%)}
& 0.1540 {\footnotesize (+153.95\%)}
& 0.2504 {\footnotesize (+14.39\%)}
& 0.2519 {\footnotesize (+7.36\%)} \\
& HR@5
& 0.6568 {\footnotesize (-1.99\%)}
& 0.6492 {\footnotesize (-1.76\%)}
& 0.6658 {\footnotesize (+4.36\%)}
& 0.6158
& \textbf{0.6770}
& 0.1835
& 0.6479 {\footnotesize (-0.51\%)}
& 0.6583 {\footnotesize (+1.70\%)}
& 0.4105 {\footnotesize (+40.45\%)}
& \underline{0.6714} {\footnotesize (+6.20\%)}
& 0.6481 {\footnotesize (+2.74\%)} \\
& HR@10
& 0.7759 {\footnotesize (-1.35\%)}
& 0.7967 {\footnotesize (+2.46\%)}
& 0.7876 {\footnotesize (+4.14\%)}
& 0.7638
& \underline{0.7993}
& 0.3250
& 0.7611 {\footnotesize (-1.82\%)}
& 0.7809 {\footnotesize (+0.91\%)}
& 0.5785 {\footnotesize (+10.14\%)}
& \textbf{0.8049} {\footnotesize (+7.83\%)}
& 0.7787 {\footnotesize (+3.80\%)} \\
& NDCG@5
& 0.4556 {\footnotesize (-4.95\%)}
& 0.4466 {\footnotesize (-4.60\%)}
& 0.4699 {\footnotesize (+5.23\%)}
& 0.4309
& 0.4721
& 0.1113
& 0.4628 {\footnotesize (+1.01\%)}
& \textbf{0.4777} {\footnotesize (+5.68\%)}
& 0.2835 {\footnotesize (+62.84\%)}
& \underline{0.4737} {\footnotesize (+8.51\%)}
& 0.4616 {\footnotesize (+3.93\%)} \\
& NDCG@10
& 0.4953 {\footnotesize (-4.22\%)}
& 0.4945 {\footnotesize (-2.29\%)}
& 0.5094 {\footnotesize (+5.06\%)}
& 0.4786
& 0.5118
& 0.1568
& 0.4998 {\footnotesize (+0.24\%)}
& \textbf{0.5176} {\footnotesize (+4.95\%)}
& 0.3377 {\footnotesize (+35.71\%)}
& \underline{0.5172} {\footnotesize (+9.18\%)}
& 0.5041 {\footnotesize (+4.26\%)} \\
\bottomrule
\end{tabular}
}
\vspace{0.5ex}
\begin{minipage}{\textwidth}
\footnotesize
\textsuperscript{\textdagger}\textsc{DHMAE} follows its original all-item evaluation protocol, whereas the other methods use the sampled evaluation protocol with one positive item and 100 negative items. Because its candidate set differs from the others, \textsc{DHMAE} is excluded from best/second-best highlighting and should be interpreted separately in cross-method comparisons.
\end{minipage}
\end{table*}

\subsection{Revisiting Performance After Correction}
\label{sec:main_findings_corrected}

Table~\ref{tab:corrected_tieaware_results} reports the corrected tie-aware results averaged over three random seeds after removing the additional sigmoid before the BPR objective, together with the relative change from the original tie-aware score. Beyond the five representative recent methods, we further examined the baselines released in the \textsc{ConsRec} codebase and found that \textsc{AGREE}, \textsc{HyperGroup}, and \textsc{HCR} are affected by the same issue and are therefore included in the corrected re-evaluation. The other baselines do not use this additional sigmoid and thus remain unchanged. \textsc{ConsRec} on \textsc{CAMRa2011} also remains unchanged because the original paper uses a dot-product scorer rather than the MLP variant.

Overall, correcting this design improves the tie-aware performance of many affected methods, confirming that the severe performance degradation observed under the original implementations is closely related to the score compression induced by the additional sigmoid. However, the magnitude of improvement varies substantially across methods and settings, and a few results even decrease slightly after correction. More importantly, after correction, strong baselines remain highly competitive across datasets and tasks, and recent methods no longer show a clear and consistent advantage. This is particularly evident in the group recommendation task, where baseline methods frequently outperform recent methods under corrected tie-aware evaluation. Taken together, these corrected results weaken the empirical case for consistent progress over strong baselines within the examined family of recent group recommendation methods.

\section{Mechanism and Mitigation}

The analyses above identify the additional sigmoid before BPR as a central source of tie inflation and evaluation bias. However, the corrected results also show that removing it does not always improve tie-aware performance. This suggests that the additional sigmoid may play a dual role: while it distorts evaluation by increasing tied scores, it may also provide a beneficial smoothing effect during optimization. We therefore examine this possibility more closely and explore a simple mitigation that decouples the potential training-side benefit from the evaluation-side distortion.

\subsection{Implicit Margin Smoothing}
\label{sec:implicit_margin_smoothing}

The results in Section~\ref{sec:main_findings_corrected} show that removing the additional sigmoid before BPR does not uniformly improve performance. In particular, for \textsc{ConsRec} on \textsc{Mafengwo} under the group recommendation task, the corrected tie-aware results are consistently worse than the original tie-aware ones across all five metrics. This suggests that the additional sigmoid is not merely an implementation artifact, but may also provide a beneficial optimization effect in some settings.

A natural interpretation is that the extra sigmoid acts as an implicit form of \emph{margin smoothing}. By squashing item scores into a bounded range before the pairwise loss is computed, it reduces the effective margin scale seen by the loss. As a result, optimization may become less sensitive to large raw-score margins and more stable during training. This effect may be particularly important for \textsc{ConsRec}. Since \textsc{ConsRec} uses a relatively lightweight shared predictor on top of a more complex fused representation, constraining the score scale may incidentally make optimization easier. At the same time, this potential benefit is entangled with a clear drawback: the same transformation can also reduce score separation at inference time, increasing the prevalence of tied scores and making top-$K$ evaluation more sensitive to tie-breaking.

These observations suggest that the additional sigmoid may simultaneously introduce a \emph{training-side smoothing effect} and an \emph{evaluation-side distortion}. The key issue is therefore not whether the transformation helps or hurts overall, but that any optimization benefit it provides is coupled with an undesirable and potentially severe evaluation bias.

\subsection{Decoupling Smoothing from Distortion}
\label{sec:decoupling_smoothing_distortion}

\begin{table}[t]
\centering
\caption{
Tie-aware results of three \textsc{ConsRec} training variants on \textsc{Mafengwo}, averaged over three random seeds: the original implementation, the variant without the additional sigmoid, and temperature-scaled BPR with $\tau=64$. Relative changes are computed against the original tie-aware score.
}
\label{tab:consrec_mafengwo_temp_scaled}
\resizebox{\columnwidth}{!}{
\begin{tabular}{lccccc}
\toprule
\multirow{2}{*}{Metric}
& Original
& \multicolumn{2}{c}{w/o additional sigmoid}
& \multicolumn{2}{c}{Temperature-scaled BPR} \\
\cmidrule(lr){2-2} \cmidrule(lr){3-4} \cmidrule(l){5-6}
& Tie-aware Score
& Tie-aware Score
& $\Delta$ vs. Orig.
& Tie-aware Score
& $\Delta$ vs. Orig. \\
\midrule
HR@1    & 0.6152 & 0.4901 & -20.33\% & 0.6221 & +1.12\% \\
HR@5    & 0.8677 & 0.7531 & -13.20\% & 0.8521 & -1.79\% \\
HR@10   & 0.8907 & 0.8270 & -7.16\%  & 0.8936 & +0.33\% \\
NDCG@5  & 0.7575 & 0.6264 & -17.31\% & 0.7512 & -0.84\% \\
NDCG@10 & 0.7650 & 0.6570 & -14.12\% & 0.7648 & -0.02\% \\
\bottomrule
\end{tabular}
}
\end{table}

The results above suggest that the additional sigmoid may provide a useful smoothing effect during optimization, meaning that removing it may not be the best remedy. We therefore consider an alternative that decouples this smoothing effect from the score distortion introduced by the additional sigmoid. Specifically, we introduce \emph{temperature-scaled BPR}, which smooths the pairwise margin directly at the loss level. Temperature scaling has been widely used to control the softness or sharpness of learning signals, for example in knowledge distillation and contrastive learning~\citep{Distilling2015,simclr2020}. We apply the same principle to the BPR margin:
\begin{equation}
\mathcal{L}_{\tau\text{-BPR}}
=
- \log \sigma \left( \frac{s_{x,i^+} - s_{x,i^-}}{\tau} \right),
\end{equation}
where $\tau > 1$ controls the degree of smoothing.

We evaluate this variant on \textsc{ConsRec} for group recommendation on \textsc{Mafengwo} over three random seeds, where removing the additional sigmoid noticeably degrades tie-aware performance. We consider $\tau \in \{1,2,4,8,16,32,64\}$ as a diagnostic sweep to characterize how increasing margin smoothing affects performance, rather than to select an optimal temperature on the test set. As shown in Table~\ref{tab:consrec_mafengwo_temp_scaled}, simply removing the additional sigmoid leads to consistent declines across all five metrics, providing evidence that the original design offers a beneficial smoothing effect during optimization in this setting. In contrast, temperature-scaled BPR with $\tau=64$, the strongest smoothing setting considered in our sweep, recovers most of this loss and even slightly surpasses the original implementation on HR@1 and HR@10, while remaining very close on the other metrics. Notably, this recovery is obtained with almost no tied scores, which supports the view that the gain comes primarily from smoothing the pairwise objective rather than from evaluation artifacts induced by tie inflation.

\begin{figure}[t]
    \centering
    \includegraphics[width=0.95\columnwidth]{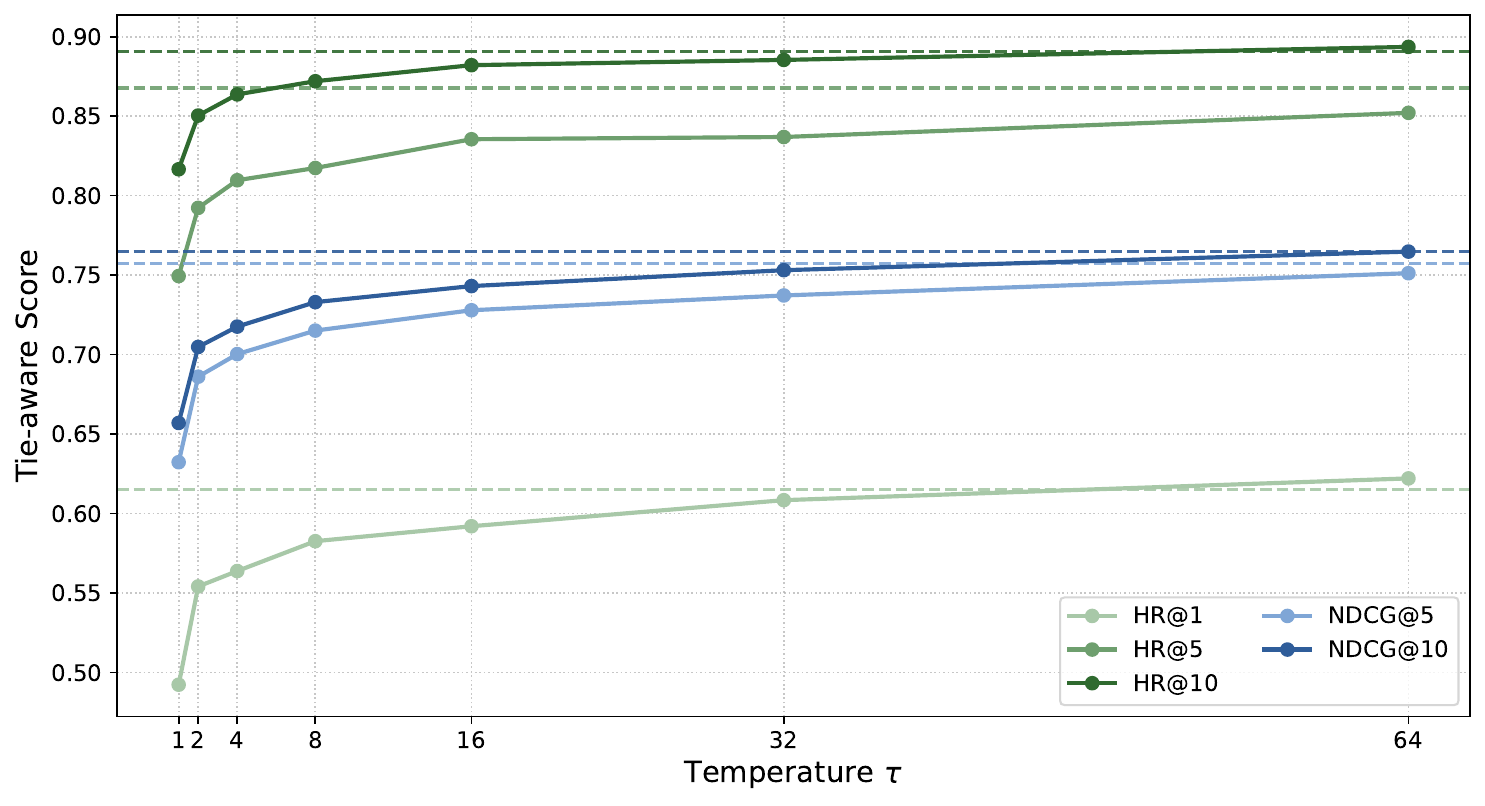}
    \caption{
    Tie-aware performance of temperature-scaled BPR on \textsc{ConsRec} over different temperatures $\tau$ on \textsc{Mafengwo}. 
    Solid lines show temperature-scaled BPR, while dashed lines mark the tie-aware scores of the original implementation.
    }
    \Description{Line plots showing the tie-aware HR@1, HR@5, HR@10, NDCG@5, and NDCG@10 of ConsRec with temperature-scaled BPR as the temperature increases. Performance generally improves with larger temperatures and approaches the tie-aware performance of the original implementation, indicated by dashed reference lines.}
    \label{fig:temp_scaled}
\end{figure}

Figure~\ref{fig:temp_scaled} further shows that tie-aware performance generally improves as the temperature increases. Across all five metrics, larger $\tau$ values consistently narrow the gap between the variant without the additional sigmoid and the original implementation, and when $\tau$ is sufficiently large, the temperature-scaled variant nearly recovers the original tie-aware performance. These results suggest that the benefit of the additional sigmoid is better understood as \emph{margin smoothing} during optimization, rather than score squashing itself.

Overall, these results indicate that the training-side benefit and the evaluation-side distortion can be partially decoupled. A simple modification to the BPR objective can retain most of the optimization benefit without directly inducing the score compression that gives rise to frequent tied scores and the resulting evaluation bias.

\section{Discussion}

Our findings suggest that, in group recommendation, evaluation design choices that are often treated as minor implementation details can materially affect empirical conclusions. When tied scores are common, reporting a single deterministic top-$K$ score can obscure substantial uncertainty in ranking quality. In this sense, the issue identified in this paper is not only a reproducibility problem tied to specific codebases, but also a broader evaluation problem: the same model predictions can yield substantially different HR@$K$ and NDCG@$K$ values under equally valid tie-breaking rules. As a result, some reported gains in recent group recommendation literature may reflect evaluation conventions rather than genuine improvements in modeling group preferences.

Tie-aware evaluation is not intended to impose a new preference over how tied items should be ordered. Instead, it makes explicit the uncertainty that already exists when the model assigns identical scores to multiple candidates. In single-positive top-$K$ evaluation, a deterministic tie-breaking rule resolves this uncertainty by selecting one feasible rank for the positive item, often according to candidate order or sorting implementation details. By averaging over all feasible ranks in the positive item's tied block, tie-aware evaluation provides an implementation-independent estimate of ranking performance whenever the model scores alone do not determine a unique ordering.

At the same time, our results should be interpreted with appropriate scope. We do not claim that all progress in group recommendation is illusory, nor that all recent methods suffer equally from the same problem. The strongest evidence in this study concerns the examined family of implementations and settings in which tied scores are prevalent, especially under single-positive top-$K$ evaluation. Moreover, although our main results are averaged over three random seeds, our corrected re-evaluations are intended primarily as diagnostic evidence for isolating the effect of score compression, rather than as an exhaustive benchmark over all possible hyperparameter configurations and model variants. Extending tie-aware reassessment to more datasets, model families, and evaluation protocols remains an important direction for future work.

These findings lead to several practical recommendations for future work. First, when tied scores are present, authors should explicitly report the tie-breaking rule used by their evaluation pipeline. Second, standard top-$K$ metrics should be accompanied by tie-aware estimates and simple tie statistics, such as top-score tie frequency and average top-score tie size. Third, training-time transformations should be evaluated not only by their effects on final accuracy, but also by how they shape the score distribution at inference time. A transformation that stabilizes optimization may simultaneously compress scores, increase tie prevalence, and make evaluation more dependent on deterministic implementation choices. Reliable progress in group recommendation therefore requires assessing ranking accuracy and tie sensitivity together.

\section{Conclusion}

We revisited recent group recommendation methods from the perspective of evaluation reliability and showed that their reported improvements can be substantially affected by an overlooked interaction between training-time score compression and evaluation-time deterministic tie-breaking. When tied scores are prevalent, standard top-$K$ metrics can produce overly optimistic and unstable estimates that do not faithfully reflect true ranking quality. Our tie-aware evaluation shows that many of the gains reported by the examined methods shrink considerably once this bias is accounted for, and that the relative ordering of methods can change markedly.

Our analysis further suggests that the additional sigmoid before BPR may provide a training-side smoothing effect while introducing evaluation-side distortion. A simple alternative, temperature-scaled BPR, can recover much of this optimization benefit without inducing severe tie inflation. Overall, our findings highlight the need for more principled evaluation in group recommendation. To support reliable progress, future work should report tie-aware metrics and tie statistics whenever tied scores are prevalent.

\begin{acks}
We thank the anonymous reviewers for their insightful and constructive feedback.
This work was supported by the National Science and Technology Council (NSTC), Taiwan, under Grant NSTC 115-2634-F-001-006.
\end{acks}
\bibliographystyle{ACM-Reference-Format}
\bibliography{references}

@inproceedings{consrec2023,
author = {Wu, Xixi and Xiong, Yun and Zhang, Yao and Jiao, Yizhu and Zhang, Jiawei and Zhu, Yangyong and Yu, Philip S.},
title = {ConsRec: Learning Consensus Behind Interactions for Group Recommendation},
year = {2023},
isbn = {9781450394161},
publisher = {Association for Computing Machinery},
address = {New York, NY, USA},
url = {https://doi.org/10.1145/3543507.3583277},
doi = {10.1145/3543507.3583277},
booktitle = {Proceedings of the ACM Web Conference 2023},
pages = {240–250},
numpages = {11},
location = {Austin, TX, USA},
series = {WWW '23}
}

@inproceedings{aligngroup2024,
author = {Xu, Jinfeng and Chen, Zheyu and Li, Jinze and Yang, Shuo and Wang, Hewei and Ngai, Edith C. H.},
title = {AlignGroup: Learning and Aligning Group Consensus with Member Preferences for Group Recommendation},
year = {2024},
isbn = {9798400704369},
publisher = {Association for Computing Machinery},
address = {New York, NY, USA},
url = {https://doi.org/10.1145/3627673.3679697},
doi = {10.1145/3627673.3679697},
booktitle = {Proceedings of the 33rd ACM International Conference on Information and Knowledge Management},
pages = {2682–2691},
numpages = {10},
location = {Boise, ID, USA},
series = {CIKM '24}
}

@inproceedings{dhmae2024,
author = {Zhao, Yingqi and Zhang, Haiwei and Bai, Qijie and Nie, Changli and Yuan, Xiaojie},
title = {DHMAE: A Disentangled Hypergraph Masked Autoencoder for Group Recommendation},
year = {2024},
isbn = {9798400704314},
publisher = {Association for Computing Machinery},
address = {New York, NY, USA},
url = {https://doi.org/10.1145/3626772.3657699},
doi = {10.1145/3626772.3657699},
booktitle = {Proceedings of the 47th International ACM SIGIR Conference on Research and Development in Information Retrieval},
pages = {914–923},
numpages = {10},
location = {Washington DC, USA},
series = {SIGIR '24}
}

@inproceedings{itr2024,
author = {Liu, Yue and Zhu, Shihao and Yang, Tianyuan and Ma, Jian and Zhong, Wenliang},
title = {Identify then recommend: towards unsupervised group recommendation},
year = {2024},
isbn = {9798331314385},
publisher = {Curran Associates Inc.},
address = {Red Hook, NY, USA},
booktitle = {Proceedings of the 38th International Conference on Neural Information Processing Systems},
articleno = {3045},
numpages = {26},
location = {Vancouver, BC, Canada},
series = {NIPS '24}
}

@article{dggvae2026,
author = {Xu, Jinfeng and Chen, Zheyu and Li, Jinze and Yang, Shuo and Wang, Wei and Wang, Hewei and Li, Yijie and Hu, Xiping and Ngai, Edith},
title = {DGGVAE: Dual-Granularity Graph Variational Auto-Encoder for Group Recommendation},
year = {2026},
issue_date = {February 2026},
publisher = {Association for Computing Machinery},
address = {New York, NY, USA},
volume = {44},
number = {2},
issn = {1046-8188},
url = {https://doi.org/10.1145/3785145},
doi = {10.1145/3785145},
journal = {ACM Trans. Inf. Syst.},
month = jan,
articleno = {52},
numpages = {30}
}

@inproceedings{camra2011,
author = {Said, Alan and Berkovsky, Shlomo and De Luca, Ernesto William and Hermanns, Jannis},
title = {Challenge on context-aware movie recommendation: CAMRa2011},
year = {2011},
isbn = {9781450306836},
publisher = {Association for Computing Machinery},
address = {New York, NY, USA},
url = {https://doi.org/10.1145/2043932.2044015},
doi = {10.1145/2043932.2044015},
booktitle = {Proceedings of the Fifth ACM Conference on Recommender Systems},
pages = {385–386},
numpages = {2},
location = {Chicago, Illinois, USA},
series = {RecSys '11}
}

@inproceedings{cao2018attentive,
author = {Cao, Da and He, Xiangnan and Miao, Lianhai and An, Yahui and Yang, Chao and Hong, Richang},
title = {Attentive Group Recommendation},
year = {2018},
isbn = {9781450356572},
publisher = {Association for Computing Machinery},
address = {New York, NY, USA},
url = {https://doi.org/10.1145/3209978.3209998},
doi = {10.1145/3209978.3209998},
booktitle = {The 41st International ACM SIGIR Conference on Research \& Development in Information Retrieval},
pages = {645–654},
numpages = {10},
location = {Ann Arbor, MI, USA},
series = {SIGIR '18}
}

@book{felfernig2018group,
author = {Felfernig, Alexander and Boratto, Ludovico and Stettinger, Martin and Tkali, Marko},
title = {Group Recommender Systems: An Introduction},
year = {2018},
isbn = {3319750666},
publisher = {Springer Publishing Company, Incorporated},
edition = {1st},
}

@ARTICLE{yin2019social,
  author={Cao, Da and He, Xiangnan and Miao, Lianhai and Xiao, Guangyi and Chen, Hao and Xu, Jiao},
  journal={IEEE Transactions on Knowledge and Data Engineering}, 
  title={Social-Enhanced Attentive Group Recommendation}, 
  year={2021},
  volume={33},
  number={3},
  pages={1195-1209},
  doi={10.1109/TKDE.2019.2936475}
}

@incollection{trattner2018evaluating,
   author = {Trattner, Christoph and Said, Alan and Boratto, Ludovico and Felfernig, Alexander},
   booktitle = {Group Recommender Systems : An Introduction},
   institution = {University of Skövde, School of Informatics},
   institution = {University of Skövde, The Informatics Research Centre},
   institution = {University of Bergen, Norway},
   institution = {EURECAT, Spain},
   institution = {Graz University of Technology, Austria},
   pages = {59--71},
   title = {Evaluating Group Recommender Systems},
   URL = {http://www.springer.com/us/book/9783319750668},
   ISBN = {978-3-319-75067-5},
   ISBN = {978-3-319-75066-8},
   year = {2018}
}

@inproceedings{rendle2009bpr,
author = {Rendle, Steffen and Freudenthaler, Christoph and Gantner, Zeno and Schmidt-Thieme, Lars},
title = {BPR: Bayesian personalized ranking from implicit feedback},
year = {2009},
isbn = {9780974903958},
publisher = {AUAI Press},
address = {Arlington, Virginia, USA},
booktitle = {Proceedings of the Twenty-Fifth Conference on Uncertainty in Artificial Intelligence},
pages = {452–461},
numpages = {10},
location = {Montreal, Quebec, Canada},
series = {UAI '09}
}

@inproceedings{dacrema2019progress,
author = {Ferrari Dacrema, Maurizio and Cremonesi, Paolo and Jannach, Dietmar},
title = {Are we really making much progress? A worrying analysis of recent neural recommendation approaches},
year = {2019},
isbn = {9781450362436},
publisher = {Association for Computing Machinery},
address = {New York, NY, USA},
url = {https://doi.org/10.1145/3298689.3347058},
doi = {10.1145/3298689.3347058},
booktitle = {Proceedings of the 13th ACM Conference on Recommender Systems},
pages = {101–109},
numpages = {9},
location = {Copenhagen, Denmark},
series = {RecSys '19}
}

@inproceedings{mcsherry2008tiedscores,
author = {McSherry, Frank and Najork, Marc},
title = {Computing information retrieval performance measures efficiently in the presence of tied scores},
year = {2008},
isbn = {3540786457},
publisher = {Springer-Verlag},
address = {Berlin, Heidelberg},
booktitle = {Proceedings of the IR Research, 30th European Conference on Advances in Information Retrieval},
pages = {414–421},
numpages = {8},
location = {Glasgow, UK},
series = {ECIR'08}
}

@article{amer-yahia2009,
author = {Amer-Yahia, Sihem and Roy, Senjuti Basu and Chawlat, Ashish and Das, Gautam and Yu, Cong},
title = {Group recommendation: semantics and efficiency},
year = {2009},
issue_date = {August 2009},
publisher = {VLDB Endowment},
volume = {2},
number = {1},
issn = {2150-8097},
url = {https://doi.org/10.14778/1687627.1687713},
doi = {10.14778/1687627.1687713},
journal = {Proc. VLDB Endow.},
month = aug,
pages = {754–765},
numpages = {12}
}

@inproceedings{dinoia2022topn,
author = {Anelli, Vito Walter and Bellog\'{\i}n, Alejandro and Di Noia, Tommaso and Jannach, Dietmar and Pomo, Claudio},
title = {Top-N Recommendation Algorithms: A Quest for the State-of-the-Art},
year = {2022},
isbn = {9781450392075},
publisher = {Association for Computing Machinery},
address = {New York, NY, USA},
url = {https://doi.org/10.1145/3503252.3531292},
doi = {10.1145/3503252.3531292},
booktitle = {Proceedings of the 30th ACM Conference on User Modeling, Adaptation and Personalization},
pages = {121–131},
numpages = {11},
location = {Barcelona, Spain},
series = {UMAP '22}
}

@article{dacrema2021troubling,
author = {Ferrari Dacrema, Maurizio and Boglio, Simone and Cremonesi, Paolo and Jannach, Dietmar},
title = {A Troubling Analysis of Reproducibility and Progress in Recommender Systems Research},
year = {2021},
issue_date = {April 2021},
publisher = {Association for Computing Machinery},
address = {New York, NY, USA},
volume = {39},
number = {2},
issn = {1046-8188},
url = {https://doi.org/10.1145/3434185},
doi = {10.1145/3434185},
journal = {ACM Trans. Inf. Syst.},
month = jan,
articleno = {20},
numpages = {49}
}

@inproceedings{lin2019scoreties,
author = {Lin, Jimmy and Yang, Peilin},
title = {The Impact of Score Ties on Repeatability in Document Ranking},
year = {2019},
isbn = {9781450361729},
publisher = {Association for Computing Machinery},
address = {New York, NY, USA},
url = {https://doi.org/10.1145/3331184.3331339},
doi = {10.1145/3331184.3331339},
booktitle = {Proceedings of the 42nd International ACM SIGIR Conference on Research and Development in Information Retrieval},
pages = {1125–1128},
numpages = {4},
location = {Paris, France},
series = {SIGIR'19}
}

@inproceedings{saha2022ties,
author = {Saha, Sourav and Roy, Dwaipayan and Mitra, Mandar},
title = {On modifying evaluation measures to deal with ties in ranked lists},
year = {2022},
isbn = {9781450393454},
publisher = {Association for Computing Machinery},
address = {New York, NY, USA},
url = {https://doi.org/10.1145/3529372.3533291},
doi = {10.1145/3529372.3533291},
booktitle = {Proceedings of the 22nd ACM/IEEE Joint Conference on Digital Libraries},
articleno = {12},
numpages = {4},
location = {Cologne, Germany},
series = {JCDL '22}
}

@inproceedings{baltrunas2010,
author = {Baltrunas, Linas and Makcinskas, Tadas and Ricci, Francesco},
title = {Group recommendations with rank aggregation and collaborative filtering},
year = {2010},
isbn = {9781605589060},
publisher = {Association for Computing Machinery},
address = {New York, NY, USA},
url = {https://doi.org/10.1145/1864708.1864733},
doi = {10.1145/1864708.1864733},
booktitle = {Proceedings of the Fourth ACM Conference on Recommender Systems},
pages = {119–126},
numpages = {8},
location = {Barcelona, Spain},
series = {RecSys '10}
}

@inproceedings{everyonesawinner2023,
author = {Shehzad, Faisal and Jannach, Dietmar},
year = {2023},
month = {09},
pages = {652-657},
title = {Everyone’s a Winner! On Hyperparameter Tuning of Recommendation Models},
doi = {10.1145/3604915.3609488}
}

@inproceedings{cabanac2010tiebreaking,
author = {Cabanac, Guillaume and Hubert, Gilles and Boughanem, Mohand and Chrisment, Claude},
year = {2010},
month = {10},
pages = {112-123},
title = {Tie-Breaking Bias: Effect of an Uncontrolled Parameter on Information Retrieval Evaluation},
isbn = {978-3-642-15997-8},
doi = {10.1007/978-3-642-15998-5_13}
}

@article{harris2020array,
  title={Array programming with NumPy},
  author={Harris, Charles R. and Millman, K. Jarrod and van der Walt, St{\'e}fan J. and Gommers, Ralf and Virtanen, Pauli and Cournapeau, David and Wieser, Eric and Taylor, Julian and Berg, Sebastian and Smith, Nathaniel J. and others},
  journal={Nature},
  volume={585},
  number={7825},
  pages={357--362},
  year={2020}
}

@inproceedings{simclr2020,
  title = {A Simple Framework for Contrastive Learning of Visual Representations},
  author = {Chen, Ting and Kornblith, Simon and Norouzi, Mohammad and Hinton, Geoffrey},
  booktitle = {Proceedings of the 37th International Conference on Machine Learning},
  pages = {1597--1607},
  year = {2020},
  volume = {119},
  series = {Proceedings of Machine Learning Research},
  publisher = {PMLR}
}

@article{distilling2015,
  title={Distilling the Knowledge in a Neural Network},
  author={Geoffrey E. Hinton and Oriol Vinyals and Jeffrey Dean},
  journal={ArXiv},
  year={2015},
  volume={abs/1503.02531},
  url={https://api.semanticscholar.org/CorpusID:7200347}
}

\end{document}